\pdfoutput=1
\RequirePackage{ifpdf}
\ifpdf 
\documentclass[pdftex]{sigma}
\else
\documentclass{sigma}
\fi

\numberwithin{equation}{section}

\begin{document}

\allowdisplaybreaks

\renewcommand{\thefootnote}{$\star$}

\renewcommand{\PaperNumber}{002}

\FirstPageHeading

\ShortArticleName{Discretisations, Constraints and Dif\/feomorphisms in Quantum Gravity}

\ArticleName{Discretisations, Constraints and Dif\/feomorphisms\\ in Quantum Gravity\footnote{This
paper is a contribution to the Special Issue ``Loop Quantum Gravity and Cosmology''. The full collection is available at \href{http://www.emis.de/journals/SIGMA/LQGC.html}{http://www.emis.de/journals/SIGMA/LQGC.html}}}

\Author{Benjamin BAHR~$^\dag$, Rodolfo GAMBINI~$^\ddag$ and Jorge PULLIN~$^\S$}

\AuthorNameForHeading{B.~Bahr, R.~Gambini and J.~Pullin}

\Address{$^\dag$~Department of Applied Mathematics and Theoretical Physics, University of Cambridge,\\
\hphantom{$^\dag$}~Wilberforce Road, Cambridge CB3 0WA, UK}
\EmailD{\href{mailto:bab26@cam.ac.uk}{bab26@cam.ac.uk}}

\Address{$^\ddag$~Instituto de F\'{i}sica, Facultad de Ciencias, Universidad de la Rep\'{u}blica,\\
\hphantom{$^\ddag$}~Igu\'{a} 4225, CP 11400 Montevideo, Uruguay}
\EmailD{\href{mailto:rgambini@fisica.edu.uy}{rgambini@fisica.edu.uy}}

\Address{$^\S$~Department of Physics and Astronomy, Louisiana State University,\\
\hphantom{$^\S$}~Baton Rouge, LA 70803-4001, USA}
\EmailD{\href{mailto:pullin@lsu.edu}{pullin@lsu.edu}}

\ArticleDates{Received November 09, 2011, in f\/inal form December 31, 2011; Published online January 08, 2012}

\Abstract{In this review we discuss the interplay between discretization, constraint implementation, and dif\/feomorphism symmetry in Loop Quantum Gravity and Spin Foam models. To this end we review the Consistent Discretizations approach, which is an application of the master constraint program to construct the physical Hilbert space of the canonical theory, as well as the Perfect Actions approach, which aims at f\/inding a path integral measure with the correct symmetry behavior under dif\/feomorphisms.}

\Keywords{quantum gravity; dif\/feomorphisms; constraints; consistent discretizations; gauge symmetries; perfect actions; renormalization}

\Classification{37M99; 70H45; 81T17; 82B28; 83C27; 83C45}  

\renewcommand{\thefootnote}{\arabic{footnote}}
\setcounter{footnote}{0}

\section{Introduction}

The dynamics of General Relativity is governed by the Einstein--Hilbert action (together with the Gibbons--Hawking--York boundary term)
\begin{gather}\label{Gl:EHAction}
S_\text{EH}[g_{\mu\nu}] = \frac{1}{8\pi}\int_Md^4x\,\sqrt{|\det g|}\left(\Lambda - \frac{1}{2} R \right) - \frac{1}{8\pi}\int_{\partial M}d^3x\,\sqrt{\det h} K .
\end{gather}
The invariance of the action (\ref{Gl:EHAction}) under space-time dif\/feomorphisms (leaving the boundary invariant) implies a large redundancy on the set of solutions, so that two solutions are to be physically equivalent if they dif\/fer by such a dif\/feomorphism (Einstein's hole argument~\cite{Wald:1984rg}). The dif\/feomorphism group therefore arises as gauge symmetry group of the theory.

Furthermore, unlike for usual gauge theories of connections where gauge transformations arise as local bundle automorphisms, the dif\/feomorphism symmetry is deeply intertwined with the dynamics of~GR: The action (\ref{Gl:EHAction}) is the only dif\/feomorphism-invariant action of a metric which leads to second order equations of motion~\cite{kuchar}. The symmetry requirement and the prescribed degrees of freedom therefore completely determine the theory.

On the canonical level  the presence of a gauge symmetry results in f\/irst class constraints, forming the so-called Dirac algebra. The Hamiltonian is a linear combination of constraints, which generate both the dynamics and space-time dif\/feomorphisms~\cite{Bergmann:1972ud}.

The two main routes for quantizing GR are either taken via the path-integral formalism, where the transition amplitudes between states $|h_{ab}\rangle$ on initial and f\/inal Cauchy surfaces $\Sigma^{i,f}$ is given by\footnote{The case in which $\Sigma^i$ is empty is called the ``no-boundary proposal'' and in this case (\ref{Gl:PathIntegralForGR}) is called the ``wave-function of the universe''~\cite{Hartle:1983ai}.}
\begin{gather}\label{Gl:PathIntegralForGR}
\langle h^f_{ab}| h^i_{cd}\rangle = \int_{g\big|\Sigma^{i,f}=h^{i,f}}\mathcal{D}g_{\mu\nu} e^{\frac{i}{\hbar}S_{\rm EH}[g_{\mu\nu}]} .
\end{gather}
Canonical quantization of GR \cite{Dirac:1958sc} on the other hand relies on a def\/inition of a state space for the boundary metrics $|\psi\rangle$ whose evolution is governed by the Wheeler--DeWitt-equation \cite{DeWitt:1967yk}
\begin{gather}\label{Gl:WdWEqn}
\hat H |\psi\rangle = 0 .
\end{gather}
Demanding that the gauge symmetry of (\ref{Gl:EHAction}) is also realized in the quantum theory either manifests itself in the condition that the path integral measure in (\ref{Gl:PathIntegralForGR}) is also invariant under dif\/feomorphisms, or that the Dirac algebra is faithfully represented on the state space of the canonical quantum theory.
Attempts to make sense of either~(\ref{Gl:PathIntegralForGR}) or~(\ref{Gl:WdWEqn}) very often rely on discretization of space-time on one way or the other, and since the interplay of discretization and space-time dif\/feomorphisms is highly nontrivial, this has been a central, and up to today unsolved issue in either attempt of formulating a theory of quantum gravity.

Out of the various proposals for a quantum gravity theory based on discretizations\footnote{To this end, see \cite{lollreview}, also in particular the Causal Dynamical Triangulations approach~\cite{Ambjorn:2010rx}, Causal Set Theory~\cite{Dowker:2006sb}, and Quantum Graphity~\cite{Konopka:2006hu}.}, we concentrate in this article on Loop Quantum Gravity \cite{LQG} and the closely related Spin Foam approach (see \cite{SPINFOAMS} for a review and literature), which can be seen as the path-integral version of canonical LQG (see e.g.~the discussion in \cite{CONNECTION}).

\subsection{Canonical approach}
LQG is a canonical approach, in which the kinematical Hilbert space is well-understood, and the states of which can be written as a generalization of Penrose's Spin-Network Functions \cite{Rovelli:1995ac}. Although the formalism is inherently continuous, the states carry many discrete features of lattice gauge theories, which rests on the fact that one demands Wilson lines to become observables. The constraints separate into (spatial) dif\/feomorphism- and Hamiltonian constraints. While the f\/inite action of the spatial dif\/feomorphisms can be naturally def\/ined as unitary operators, the Hamiltonian constraints exist as quantizations of approximate expressions of the continuum constraints, regularized on an irregular lattice. It is known that the regularized constraints do not close, so that the algebra contains anomalies, whenever the operators are def\/ined on f\/ixed graphs which are not changed by the action of the operators \cite{wp}. If def\/ined in a graph-changing way, the commutator is well-def\/ined, even in the limit of the regulator going to zero in a controlled way \cite{Thiemann:1996aw}. However, the choice of operator topologies to choose from in order for the limit to exist is nontrivial \cite{Lewandowski:1997ba}, and the resulting Hamiltonian operators commute \cite{Thiemann:1998xy}. Since they commute on dif\/feomorphism-invariant states, the constraint algebra is satisf\/ied in that sense. Furthermore, however, the discretization itself is not unique, and the resulting ambiguities survive the continuum limit \cite{Perez:2005fn}. In the light of this, it is non-trivial to check whether the correct physics is encoded in the constraints.

Generically, the interaction of discretizations and constraints is intricate. Physical theories based on continuous variables are typically
discretized for two dif\/ferent purposes: a) to turn the dif\/ferential
equations of the theory into dif\/ference equations in order to solve
them numerically; b) to regularize and quantize the theory, as in
lattice gauge theory. When the equations of the theory describe a free
time evolution the main concern when discretizing is to achieve long
term numerical stability. When the theories have constraints among
their variables in addition to evolution equations the situation
complicates. Typically the constraints satisfy certain conditions.
Examples of such conditions are constraints that structure themselves
into Poisson algebras in canonically formulated theories. Moreover
constraints have to hold for all times, which implies that they need
to be preserved by the evolution equations. The problem is that if one
simply discretizes constraints and evolution equations these
conditions generically get violated. At an operational level, to prove
that constraints form an algebra or that they are preserved upon
evolution one uses repeatedly the Leibniz property for derivatives,
which fails to hold for discrete versions of derivatives. At a more
profound level, constraints are associated with symmetries of theories
that generically get broken by the discretization. If one insists in
using the discrete equations, one is faced with the fact that
generically they are inconsistent: they cannot be solved
simultaneously. This phenomenon is well known in numerical relativity,
where people proceed by ignoring some of the equations and solving the
rest in the hope that the ignored equations, when evaluated on the
solution, will be non-zero but small. Some algorithms can actually
achieve this for long enough evolutions and that is the state of the
art of that f\/ield.

These problems are pervasive even in simple theories. People tend to
get misled by the success of lattice gauge theory without realizing
that a careful choice of staggered discretization is used in order to
preserve Gauss' law. In fact one of the central properties of lattice
gauge theories is that it provides a gauge invariant regularization,
i.e. it does not break the symmetries of the theory. To give some
perspective, it is worthwhile noting that even a relatively simple
theory like Maxwell theory has to be discretized judiciously. Attempts
to integrate Maxwell's equations numerically ran into signif\/icant
problems until Yee~\cite{yee}, in a landmark paper, introduced a~staggered discretization that preserves Gauss' law.

Going to the case of interest, that of general relativity and other
dif\/feomorphism invariant theory, we currently do not know of any
discretization scheme that would preserve the constraints of the
theory. So we are, ef\/fectively, where people were with respect to
Maxwell's theory before Yee's algorithm. Attempts have been made to
def\/ine a discrete calculus, but up to present successful
implementations only are possible in the linearized theory.

In Section~\ref{Ch:Consistent}, we will discuss these issues in more details for models that share important features with General Relativity, and demonstrate the consistent and uniform discretization approach, in order to deal with the aforementioned problems.

\subsection{Covariant approach}

On the covariant level, a very convenient and geometric way of discretizing GR is Regge Calculus, in which the metric variables are discretized on a triangulation, i.e.~a separation of space-time into discrete building blocks. Although evolved from a dif\/ferent perspective, the Spin Foam approaches can be seen as providing a path integral for Regge Calculus \cite{Regge:1961px} in connection-, rather than metric variables.

It is known that, generically, discretizing mechanical systems which possess a repara\-met\-ri\-za\-tion-invariance breaks this invariance \cite{marsdenwest} (see also the discussion in \cite{Bahr:2009qc}). There has been an intensive discussion in the literature about whether some invariance akin to dif\/feomorphisms exist in the case of Regge Calculus \cite{HERBIEBIANCHI,Dittrich01}. While it is well-known that some very specif\/ic solutions, describing a f\/lat metric, can be related via continuous gauge transformations \cite{Hamber:1992df,Rocek,Dittrich:2007wm}, it was also demonstrated in \cite{bahrdittrich1} that this symmetry is broken whenever solutions describe geometries with curvature, and the breaking mechanism is completely analogous to what happens in mechanical systems. Since the current Spin Foam models, at least in the semiclassical limit of large quantum numbers, reproduce Regge Calculus \cite{Barrett:2010ex}, it has to be expected that the same breaking of dif\/feomorphism invariance also appears in the path integral measure.

In Section \ref{Sec:ImprovedAndPerfectActions} we will discuss the mechanism of how dif\/feomorphism-symmetry is broken in Regge Calculus, and describe an attempt to regain it by ``improving'' the Regge action. We will show how one can attempt to construct a discrete action with the correct symmetries by a coarse graining procedure, resembling a Wilsonian renormalization group f\/low. We will also demonstrate how a similar procedure works for the toy example of the (an-)harmonic oscillator, closing the section with remarks on consequences for renormalization in Spin Foam models. Also, it will be argued how the symmetry breaking the path integral measure will account for divergencies in the ``sum over triangulations''.

Although LQG and Spin Foam models agree well on the kinematical level, the connection of their dynamics is still not understood perfectly \cite{CONNECTION, Alexandrov:2010un}. Therefore it is hard to argue that the methods described in this article exactly correspond to each other. Still, they are both attempts to solve problems that arise due to the introduction of a discretization. Also, in \cite{hohn} it was shown that, for a theory of gravity discretized on a triangulation, f\/inding an action with the right symmetries, and f\/inding constraints satisfying the Dirac algebra  (and therefore generating deformations of the triangulated Cauchy surface), are equivalent. For a toy model, this was shown to be true for the quantum case, i.e.~a path integral measure with the right kind of symmetries, and the propagator being the correct projector onto the physical Hilbert space, this was shown in \cite{Bahr:2011uj}. In this sense, the problems on the canonical and the covariant side are directly related to each other, and the two approaches described in this article can be viewed as addressing the problem from dif\/ferent angles.

It is important to note that the breaking of gauge symmetry in the current approaches to Quantum Gravity arise due to the intricate relationship between space-time dif\/feomorphisms and the discretization involved. Although it is widely expected that any Quantum Gravity theory should predict a fundamental space-time discreteness due to the presence of a minimal length scale, this does not necessarily require a breaking of dif\/feomorphism symmetry, but is rather a statement about the discreteness of the spectrum of geometric operators measuring lengths, areas, etc.~\cite{Rovelli:2002vp}.

\section{Consistent discretisation}\label{Ch:Consistent}

Consistent discretisations \cite{consistent} were an attempt to discretise theories
like general relativity in which the resulting discretised equations
were algebraically consistent, that is, they could be solved
simultaneously. Notice that this is not what happens if one just
takes the Einstein equations and discretises them. The Einstein
equations divide into constraints that hold at every instant of time
and evolution equations that evolve the variables. It turns out that
if one evolves the variables with the discrete equations, at the end
of the evolution the variables fail to satisfy the constraints. The
resulting equations are therefore inconsistent. This is well known,
for instance, in numerical relativity. The way consistent
discretisations will make the equations consistent is by determining
the values of the Lagrange multipliers. That is, the latter stop being
free variables in the discrete theory. We will see that a problem
arises in that the equations determining the Lagrange multipliers
are polynomials and therefore it is not guaranteed that the multipliers
remain real. And since one loses control of their values, one cannot
guarantee that the resulting discretisation will approximate well the
continuum theory. This problem is addressed by the Uniform Discretisations,
which we discuss in Section~\ref{Sec:UNIFDISC}.

Our goal would be to have a canonical quantization of a system in
question using a discretization. Let us outline the steps we would
like to see carried out:
\begin{itemize}\itemsep=0pt
\item We discretize the theory to be analyzed.
\item We study the resulting discrete theory at a classical level (in
  particular making sure it is a well def\/ined theory).
\item We canonically quantize such a theory.
\item We either take the continuum limit of the resulting quantum
  theory or show that it does not exist.
\end{itemize}

We are particularly interested in studying systems with common features
to those of general relativity: invariance under dif\/feomorphisms, the
presence of f\/irst class constraints, the theory being totally
constrained, among others.

If the starting theory is invariant under dif\/feomorphisms, the
discretization inevitably will break that symmetry. The discrete
theory has less gauge symmetries and therefore more degrees of freedom
than the continuum one. When one works in a Hamiltonian picture, space
and time have asymmetrical roles, so the breakage of dif\/feomorphism
symmetry manifests itself dif\/ferently in space than in time. If one
only discretizes in time, the breaking of the invariance leads to the
corresponding constraints being absent in the discrete theory and the
corresponding Lagrange multipliers get determined. We will see an
example soon. If one discretizes spatially only, the breakage
translates itself in the constraints associated with spatial
dif\/feomorphisms becoming second class and therefore fail to be
generators of symmetries.

\subsection{Systems with discrete time}

There is an alternative way to construct a discretized version of a
theory that instead of simply discretizing evolution equations and
constraints starts by discretizing the action of the theory. One
subsequently works out the equations of motion of such action. Unless
pathologies arise, the resulting equations are consistent, as they
are simply the equations of motion of an action. We call this approach
{\em consistent discretizations} and refer the reader to
\cite{consistent} for more details. Let us discuss how
this works in the case of a simple mechanical system.

Suppose one has a mechanical system with Lagrangian
$\hat{L}(q,\dot{q})$ where generically $q$ and $\dot{q}$ are~$N$
component vectors, representing a system with $N$ degrees of freedom.
We discretize time in equal intervals
$t_{n+1}-t_n=\varepsilon$. We denote the coordinates at $t_n$ as $q_n$
For simplicity we use a~f\/irst order
approximation for derivatives $\dot{q}=(q_{n+1}-q_n)/\varepsilon$. One
can choose other approximations for the derivatives that may
perform better numerically, but here for simplicity we keep the f\/irst
order one. We def\/ine a Lagrangian for the discrete theory as
\begin{gather*}
L(n,n+1) \equiv L(q_n,q_{n+1})\equiv\varepsilon \hat{L}(q,\dot{q}).
\end{gather*}
In terms of this Lagrangian the action can be written as
\begin{gather*}
S= \sum_{n=0}^N L(q_n,q_{n+1})
\end{gather*}
and from it we can derive the equations of motion
\begin{gather}\label{eldisc}
{\partial S \over \partial q_n} = {\partial L(q_{n-1},q_{n}) \over
\partial
q_n}+{\partial L(q_{n},q_{n+1}) \over \partial q_n} =0.
\end{gather}
Notice that in the discrete theory the Lagrangian is a function of~$q_n$ at the $n$-th level and of~$q_{n+1}$ at the next level (or
equivalently $q_n$ and $q_{n-1}$).

In order to formulate the theory canonically we need to def\/ine
conjugate momenta. At f\/irst this may appear surprising. If there is no
variable $\dot{q}$ anymore, how could there be a canonical momentum?
In the continuum the Lagrangian is a function of the cotangent bundle
of the conf\/iguration space $C$. In the discrete theory the Lagrangian is a
function of $C\times C$. One can change from the coordinatization
$(q_n,q_{n+1})$ to $(q_n,p_n)$ def\/ining the momentum.
For that purpose it is useful to relate the
continuum and discrete derivatives of the Lagrangians
\begin{gather}
\frac{\partial L(n,n+1)}{\partial q_{n+1}}  =  \frac{\partial
\hat{L}}{\partial \dot{q}}, \label{der1}\\
\frac{\partial L(n,n+1)}{\partial q_n}  =
\varepsilon\frac{\partial \hat{L}}{\partial q} -\frac{\partial
\hat{L}}{\partial \dot{q}}.\label{der2}
\end{gather}
The Lagrange equations in the continuum can be written as
\begin{gather*}
\frac{d\,p}{dt} = \frac{\partial \hat{L}}{\partial q}.
\end{gather*}
To discretize this last expression we make the substitution
\begin{gather*}
\dot{p} \rightarrow \frac{p_{n+1}-p_n}{\varepsilon}
\end{gather*}
and use (\ref{der1}), (\ref{der2})  to get
\begin{gather*}
\frac{p_{n+1}-p_n}{\varepsilon} = \frac{\partial \hat{L}}{\partial
q} =\frac{1}{\varepsilon}\left[\frac{\partial \hat{L}}{\partial
\dot{q}} +\frac{\partial L(n,n+1)}{\partial q_n}\right]\\
\hphantom{\frac{p_{n+1}-p_n}{\varepsilon}}{}
=\frac{1}{\varepsilon}\left[ \frac{\partial L(n,n+1)}{\partial
q_{n+1}}  +\frac{\partial L(n,n+1)}{\partial q_n}\right] =
\frac{1}{\varepsilon}\left[ p_{n+1}  +\frac{\partial
L(n,n+1)}{\partial q_n}\right].
\end{gather*}
Then
\begin{gather*}
  p_n = - \frac{\partial
L(n,n+1)}{\partial q_n}.
\end{gather*}
The resulting discrete Lagrange equations are
\begin{gather}\label{eomdisc}
p_{n+1} = \frac{\partial L(q_n,q_{n+1})}{\partial q_{n+1}}
 ,\qquad p_n = - \frac{\partial L(q_n,q_{n+1})}{\partial q_n}
 .
\end{gather}
It is worthwhile pointing out that these equations def\/ine a type 1
canonical transformation from the variables $(q_n,p_n)$ to
$(q_{n+1},p_{n+1})$.
This transformation is
canonical, in the sense that it preserves the symplectic structure
\begin{gather*}
\left\{q_n,p_n\right\}=\left\{q_{n+1},p_{n+1}\right\}=1.
\end{gather*}

In order to make contact with the continuum theory, we point out that
if one takes as discrete Lagrangian the Jacobi principal function
$S(q,Q,\varepsilon)$ then the discrete evolution  given by
(\ref{eomdisc}) will correspond to the evolution of the continuous
system for times $t=n \varepsilon$. This is an alternative view on
discrete Lagrangians: they are approximants to the function
$S(q,Q,\varepsilon)$. This is closely related to the idea of a
symplectic (also known as variational) integrator \cite{marsdenwest}.

We note that there exists a discrete version of Noether's theorem.
If the system has a global symmetry $\delta_u q_n= u \delta q_n$, then
the discrete Lagrangian will be invariant up to a total dif\/ference
\begin{gather*}
  \delta_u L=u\left(\frac{\partial L(q_{n},q_{n+1})}
    {\partial q_n} \delta q_n
+\frac{\partial L(q_{n},q_{n+1})}{\partial q_{n+1}}
\delta q_{n+1}\right)=u(B_{n+1}-B_n),
\end{gather*}
and using the equation of motion  (\ref{eldisc}), we get the conserved
quantity{\samepage
\begin{gather*}
C_n=-\frac{\partial L}{\partial q_n} \delta q_n+B_n
\end{gather*}
(that is $C_{n+1}=C_n$ ).}

If the action is invariant under a ``local'' transformation
$\delta_u q_n= u(n) \delta q_n$, we have
\begin{gather*}
0=\delta_u S=\sum_n \frac{\partial S}{\partial q_n} \delta q_n u(n),
\end{gather*}
and since the $u(n)$
are arbitrary, this means that $\frac{\partial S}{\partial q_n} \delta
q_n=0$,
that is
\begin{gather}\label{idnoetherdisc}
\left(\frac{\partial}{\partial q_n}\left(L(q_{n-1},q_n)+L(q_{n},q_{n+1}) \right) \right)\delta q_n=0.
\end{gather}
This implies that the equations of motion (\ref{eldisc})
will not be independent and therefore one cannot unambiguously solve
 $q_{n+1}$ in terms of $q_{n}$ and $q_{n-1}$.
That is, the system will be singular. In particular the Legendre transform
$q_{n+1} \to p_n=-\frac{\partial L_\epsilon}{\partial q_n}$
will be singular and there will be constraints in phase space. The
framework to treat constrained systems of this sort parallels Dirac's
method in the continuum and is described in the
second reference of~\cite{consistent}.

Let us conclude our classical discussion considering an example of a
simple singular discrete system: a parameterized free particle whose
continuum Lagrangian is $\tilde{L}=\frac{m}{2} \dot{x}^2/\dot{t}$, and
we consider the following discretization
\begin{gather*}
L(x_n,t_n,x_{n+1},t_{n+1})=\frac{m}{2}\frac{(x_{n+1}-x_n)^2}{(t_{n+1}-t_n)}.
\end{gather*}
This discretization inherits its singular nature from the continuum.
In particular, when we compute the momenta
$p^x_n=-\partial L
/ \partial x_n$ and  $p^t_n=-\partial L / \partial t_n$ we see they are
not independent but they satisfy the same relations as the continuum ones
\begin{gather*}
\phi=p^t_n+\frac{1}{2 m} (p^x_n)^2=0.
\end{gather*}
The ``local'' invariance of the Lagrangian is given by the
transformations $\delta_u x_n=u(n)\delta x_n$, $\delta_u
t_n=u(n)\delta t_n$, where
\begin{gather*}
\delta t_n   =   1, \qquad
\delta x_n   =   \frac{1}{2 m}\left(\frac{x_{n+1}-x_{n}}{t_{n+1}-t_{n}} + \frac{x_{n}-x_{n-1}}{t_{n}-t_{n-1}}\right),
\end{gather*}
and this transformation satisf\/ies (\ref{idnoetherdisc}).

In phase space the transformation is given by
$\delta t_n = 1$ and  $\delta x_n=1/(2m)(p^{x}_n+p^{x}_{n-1})$.
Using the equations of motion we have that
$p^{x}_{n-1}=p^{x}_{n}$ which implies
\begin{gather*}
\delta t_n   =   1,\qquad
\delta x_n   =   \frac{1}{m}p^x_n .
\end{gather*}
This is the general transformation generated by the constraint
$\phi$.

Let us brief\/ly discuss how to quantize these discrete systems. The
general treatment is discussed in~\cite{norton}.
The treatment is similar to ordinary quantum mechanics, but evolving
in discrete intervals~$\varepsilon$. The starting point is as in the
continuum, one considers wavefunctions~$\psi(q)$, def\/ining the
operators $\hat{q}$ and $\hat{p}$ as usual, and the canonical
evolution is implemented through a unitary operator.
That is, if we call $Q(q,p)$ and $P(q,p)$ to the solution of the
classical equations of motion  (\ref{eomdisc}), we need a unitary
operator  $\hat{U}$ such that
\begin{gather} \label{heisdisc}
Q(\hat{q},\hat{p})   =  \hat{U} \hat{q} \hat{U}^\dag, \qquad
P(\hat{q},\hat{p})   =   \hat{U} \hat{p} \hat{U}^\dag,
\end{gather}
which represent the discrete evolution equations in the Heisenberg
representation.

In the position basis equations (\ref{heisdisc}) are a set of
dif\/ferential equations that determine the matrix $U(q',q)$. In some
cases one can show that the solution to such equations is $U(q',q)= C
\exp \frac{i}{\hbar}L_\epsilon (q,q')$ with $C$ a proportionality
constant.

We illustrate the quantization  with
a simple example given by  a particle in a potential. The Lagrangian is
\begin{gather*}
L(q_{n},q_{n+1})= m {(q_{n+1}-q_{n})^2 \over 2 \epsilon} - V(q_{n})
\epsilon
\end{gather*}
the canonical momentum is given by $p_{n+1} = m
(q_{n+1} -q_{n})/\epsilon$, from which we can get $q_{n+1}=p_{n+1} \epsilon/m
+q_{n}$. The generating function is,
\begin{gather*}
F_2=p_{n+1} q_{n} + {p_{n+1}^2 \over 2 m} \epsilon +V(q_{n+1}) \epsilon =
p_{n+1} q_{n} +H(p_{n+1}, q_{n}),
\end{gather*}
and the equations  of motion derived from the corresponding canonical
transformation are
\begin{gather*}
q_{n+1} = q_{n}+{p_{n+1} \over m} \epsilon,\qquad
p_{n}  =  p_{n+1} + V'(q_{n}) \epsilon,
\end{gather*}
which can be solved for $p_{n+1}$ as
\begin{gather*}
q_{n+1}  =  q_{n}+ {p_{n} \over m} \epsilon -V'(q_{n})
 {\epsilon^2\over m},\qquad
p_{n+1}  =  p_{n} - V'(q_{n})
\epsilon.
\end{gather*}

To quantize the system we choose a polarization such
that the wavefunctions are functions of the conf\/iguration variables,
$\Psi(q_{n+1})$. The canonical operators have the usual form. The
evolution of the system is given by a unitary
transformation, $p_{n+1} = U p_{n} U^\dagger$,
$q_{n+1}=U q_{n} U^\dagger$ and the unitary operator $U$ is given by
\begin{gather*}
U= \exp\left(i{V(q_{n})\epsilon\over \hbar}\right)
\exp\left(i{p_{n}^2\epsilon\over 2m \hbar}\right).
\end{gather*}

Quantum mechanically, the energy $H^0(q_{n+1},p_{n+1})$ is not
conserved, as expected since it was not conserved
classically. It is remarkable however, that one can construct an
``energy'' (both at a quantum  and classical level) that is
conserved by the discrete evolution. This can be accomplished using
the Baker--Campbell--Hausdorf\/f formula
\begin{gather*}
\exp\left(X\right) \exp\left(Y\right)= \exp\left(X+Y+{1 \over 2}
\left[X,Y\right]+ {1 \over 12}\left(\left[X,\left[X,Y\right]\right]+
\left[Y,\left[Y,X\right]\right]\right)+\cdots\right),
\end{gather*}
and one can therefore write $U=\exp\left({i\epsilon \over \hbar}
H_{\rm ef\/f}(q_i,p_i)\right)$ where $H_{\rm
ef\/f}=H^0(q,p)+O\left(\epsilon^2\right)$, which is
conserved under evolution. It is straightforward to write down
a classical counterpart of this expression.

Let us now return to the classical theory and turn our attention
towards systems with constraints in the continuum. In that case the
discrete Lagrangian will be{\samepage
\begin{gather*}
L(n,n+1) = p_n (q_{n+1}-q_n) -\epsilon H(q_n,p_n)-\lambda_{nB} \phi^B(q_n,p_n),
\end{gather*}
where we assume we have $M$ constraints $B=1,\ldots, M$.}

We construct the appropriate canonically conjugate
momenta using the f\/irst of (\ref{eomdisc})
\begin{gather}
P^q_{n+1}  =  {\partial L(n,n+1) \over \partial q_{n+1}} =p_n,\label{37}\\
P^p_{n+1} =  {\partial L(n,n+1) \over \partial p_{n+1}}= 0,\label{38}\\
P^{\lambda_B}_{n+1} =  {\partial L(n,n+1)
\over \partial \lambda_{(n+1)B}}= 0.\label{49}
\end{gather}
Notice that the canonically conjugate momentum of the variable $q$ at
instant $n+1$ is equal to the variable that superf\/icially appeared as
canonically conjugate at time $n$.

To determine the equations of motion for the system we start from the
second set of equations~(\ref{eomdisc})
\begin{gather*}
P^q_n  =  -{\partial L(n,n+1)
\over \partial q_{n}} =p_n+\epsilon
{\partial H(q_n,p_n) \over \partial q_n} + \lambda_{nB}
{\partial \phi^B(q_n,p_n) \over \partial q_n},\\ 
P^p_n  =  -{\partial L(n,n+1)
\over \partial p_{n}} = -(q_{n+1}-q_n) +
\epsilon {\partial H(q_n,p_n) \over \partial p_n}
+ \lambda_{nB}
{\partial \phi^B(q_n,p_n) \over \partial p_n},\\ 
P^{\lambda_B}_n  =  \phi^B(q_n,p_n).
\end{gather*}

Combining the last two sets of equations we get the equations of
motion for the system
\begin{gather}
p_n-p_{n-1}  =  -\epsilon {\partial H(q_n,p_n) \over \partial q_n} - \lambda_{nB}
{\partial \phi^B(q_n,p_n) \over \partial q_n},\label{43}\\
q_{n+1}-q_n  =
\epsilon {\partial H(q_n,p_n) \over \partial p_n}
+ \lambda_{nB} {\partial \phi^B(q_n,p_n) \over \partial p_n},\label{44}\\
\phi^B(q_n,p_n)  = 0.\label{45}
\end{gather}

These equations appear very similar to the ones one would obtain by
f\/irst working out the equations of motion in the continuum and then
discretizing them. A signif\/icant dif\/ference, however, is that when
one solves this set of equations the Lagrange multipliers get
determined, they are not free anymore as they are in the continuum.
We consider the constraint equation~(\ref{45}) and substitute~$p_n$ by~(\ref{37})
\begin{gather*}
\phi^B\big(q_n,P^q_{n+1}\big)=0.
\end{gather*}
We then solve (\ref{44}) for $q_n$ and substitute it in the previous equation,
one gets
\begin{gather*}
\phi^B\big(q_{n+1},P^q_{n+1},\lambda_{nB}\big)=0, 
\end{gather*}
and this constitutes a system of equations. Generically, these will
determine
\begin{gather*}
\lambda_{nB} =\lambda_{nB}\big(q_{n+1},P^q_{n+1}, v^\alpha\big), 
\end{gather*}
where the $v^\alpha$ are a set of free parameters that may arise if
the system of equations is undetermined. The eventual presence of
these parameters will signify that the resulting theory still has a
genuine gauge freedom represented by freely specif\/iable Lagrange
multipliers. We therefore see that generically when one discretizes
constrained theories one gets a dif\/ferent structure than in the
continuum in which Lagrange multipliers get determined.

The main advantage of this approach is that the resulting set of
equations are {\em consistent}. The equations that in the continuum
used to be constraints become upon discretization pseudo-constraints
in that they relate variables at dif\/ferent instants of time and are
solved by determining the Lagrange multipliers. The approach has been
successfully applied in various example systems~\cite{consistent}
including cosmologies~\cite{discretecosmo} and used to evolve
numerically the Gowdy space-ti\-me~\cite{gowdy}. However, in pursuing
those examples, particularly the Gowdy one which has a reasonable
level of complexity, a drawback to this approach became quite evident,
that led to a reformulation we will discuss next.

Before continuing it is worthwhile emphasizing some aspects of the
discrete theories we have been constructing:
\begin{itemize}\itemsep=0pt
\item The discrete Lagrangian is independent of the step
  $\varepsilon$.
\item The discrete system is unconstrained, or at least has less
constraints than the continuum one. If there are no constraints the
evolution equations can be solved without ambiguities.
\item Since the discrete system has more degrees of freedom than the
  continuum one there will exist trajectories that have no relation to
  those of the continuum (for instance trajectories far from the
  constraint surface).
\item The Lagrange multiplier in the discrete theory is not free but
  becomes a well def\/ined function of phase space. Its precise form
  will depend on the discretization, but in general it will vanish on
  the constraint surface. The closer the discrete evolution is to the
  constraint surface of the continuum theory, the smaller the values of
  the Lagrange multipliers.
\item It could happen that the Lagrange multipliers in the discrete
  theory become complex.
\end{itemize}

Let us assume that the discrete theory is well def\/ined. In that case
we can proceed to quantize as we described. Supposing one can f\/ind the
operator $\hat{U}$ that implements the evolution, one has a complete
description of the quantum theory. The question now is: how do we take
the continuum limit? The question is non-trivial since we don't have
at hand the parameter $\varepsilon$ that controls the continuum
limit. The classical continuum theory is achieved in the limit in
which $q_{n+1}\to q_n$, which is equivalent to taking initial
conditions close to the constraint surface. Quantum mechanically one
would like something like ``$\hat{q}_{n+1}\to \hat{q}_n$''. Since the
evolution operator is given (it does not depend on any parameter
$\varepsilon$) such a condition can only be imposed at the level of
the Hilbert space, considering states $\vert \psi \rangle$ such that
$\hat{U}\vert \psi \rangle \to \vert \psi \rangle$. We will therefore
say that the continuum theory is def\/ined by states such that
$\hat{U}\vert \psi \rangle = \vert \psi \rangle$. Later on, in the
context of uniform discretizations, we will relate this condition with
the def\/inition of physical space in the Dirac quantization of
constrained systems.

\subsection{Uniform discretizations}\label{Sec:UNIFDISC}

The main drawback of the approach presented in the previous subsection
is that the equations that determine the Lagrange multipliers are
non-linear algebraic equations. Generically their solutions are
therefore complex. Given that the Lagrange multipliers in general
relativity are the lapse and shift, they cannot be complex. That means
that the discrete theory can suddenly fail to approximate the
continuous theory in a dramatic way as it becomes complex. Also, since
the lapse and the shift determine the size of the temporal and spatial
discrete steps, one typically would not want their values to become
too large, otherwise the discrete equations will surely dif\/fer quite
signif\/icantly or may even miss features from the continuum
theory. Unfortunately one has no control on these matters. Given some
initial data, the lapse and shift will evolve dynamically. If one
encounters that they become complex, negative or too large one cannot
do anything but change the initial data. This lack of control in the
approximation clearly is a very undesirable feature, both for
numerical evolutions but also for quantization.

To address the above problem we would like to make use of the
considerable freedom that one has at the time of discretizing a
theory. There exist inf\/initely many discrete theories that approximate
a continuum one. We will use the freedom to demand that the evolution
equations of the discrete theory take the form
\begin{gather*}
A_{n+1} = e^{\{\bullet,H\}}(A_n)\equiv A_n +\{A_n,H\}
+{1 \over 2}\{\{A_n,H\},H\}+\cdots.
\end{gather*}
The quantity $H$ is def\/ined in the following way:
Consider a smooth function of $N$
variables $f(x_1,\ldots,x_N)$ such that the following three conditions
are satisf\/ied: a) $f(x_1,\ldots,x_N)=0 \iff x_i=0$ $\forall\, i$ and
otherwise $f>0$; b) ${\partial{f}\over\partial{x_i}}(0,\ldots,0)=0$;
c) $\det {\partial^2 f \over \partial x_i \partial x_j}\neq 0$
$\forall\, x$ and d) $f(\phi_1(q,p),\ldots$,
$\phi_N(q,p))$ is def\/ined for all $(q,p)$ in the complete phase space.
Given this we def\/ine $H(q,p)\equiv f(\phi_1(q,p),\ldots,
\phi_N(q,p))$.

A particularly simple example is
\begin{gather}\label{H}
H(q,p) =\frac 12 \sum_{i=1}^N \phi_i(q,p)^2,
\end{gather}
essentially the ``master constraint''
of the ``Phoenix project'' \cite{phoenix}.  The key observation is
that the above evolution preserves the value of $H$. Therefore if one
chooses initial data such that the constraints are small, they are
guaranteed to remain small upon evolution. The evolution step is
controlled by the value of $H$ and therefore by how small the
constraints are. We call this approach {\em uniform discretizations}
and the reader is referred to \cite{uniform} for more details.

At f\/irst the chosen form for the evolution may sound
counterintuitive. After all, if one gave data that satisfy the
constraints exactly, there would be no evolution. Let us show that
indeed the evolution approximates the continuum evolution well.  Let
$H$ as in the simple example above and take its initial value to be
$H_0=\delta^2/2$. We def\/ine $\lambda_i=\phi_i/\delta$, and therefore
$\sum\limits_{i=1}^N \lambda_i^2 =1$. The evolution of the dynamical variable
$q$ is given by
\begin{gather*}
q_{n+1}=q_n +\sum_{i=1}^N\{q_n,\phi_i\} \lambda_i\delta +O\big(\delta^2\big)
\end{gather*}
and if we def\/ine $\dot{q}\equiv \lim\limits_{\delta\to
0}(q_{n+1}-q_n)/\delta$, where we have identif\/ied the ``time
evolution'' step with the initial data choice for $\delta$, one then
has
\begin{gather*}
\dot{q} =\sum_{i=1}^N\{q,\phi_i\}\lambda_i,
\end{gather*}
and similarly for other dynamical variables. The particular values of
the multipliers $\lambda_i$ depend on the initial values of the
constraints $\phi_i$. Notice that as in the consistent discretizations
approach, the Lagrange multipliers get determined, but they are well
def\/ined real functions of phase space. And note that the Lagrange
multipliers are proportional to the constraints of the continuum
theory. Therefore if the evolution is kept close to the constraint
surface, the Lagrange multipliers will be small. If one chose initial
data that exactly solved the continuum constraints, the Lagrange
multipliers will vanish and the system does not evolve.

To illustrate that the above construction is actually feasible, let us
consider a one-dimensional system, the parameterized harmonic
oscillator. The continuum action for this system is
\begin{gather*}
S=\frac{1}{2}\int \left(\dot{x}^2/\dot{t}-x^2\dot{t} \right)d\tau,
\end{gather*}
where we have chosen $m=k=1$ and the potential $V(x)=k x^2/2$ as usual.
The Lagrangian that yields the discrete evolution is given by
\begin{gather*}
L_U(x,t,X,T)=S(x,X,T-t)+\frac{1}{2}(T-t)^2,
\end{gather*}
where $S$ is Hamilton's principal function of the continuous system
and we introduced the shorthand $t=t_n$, $T=t_{n+1}$, $x=x_n$ and
$X=x_{n+1}$.

The equations for $X$ and $x$ give the equations of motion for a
harmonic oscillator at a time $(T-t)$. When we work out the equations
for $T$ and $t$ we have two terms
\begin{gather*}
P_t=p_t=\frac{\partial S(x,X,T-t)}{\partial T}+ (T-t).
\end{gather*}
Since $S$ satisf\/ies by the Hamilton--Jacobi equation, we have that
\begin{gather*}
 \frac{\partial S(x,X,T-t)}{\partial T}=-H(x,p_x)=-\frac{1}{2}\left(p_x^2+x^2\right),
\end{gather*}
so the equation for $T$ is
\begin{gather*}
T=t+p_t+H(x,p_x)=t+\phi(q,p).
\end{gather*}
That is, $\lambda_d(q,p)=\phi(q,p)$ as we expected. Note that this
construction can be carried out for any parameterized system.

The constants of motion of the discrete theory are functions
$O^D$ such that $\{O^D,H\}=0$. On the constraint surface, such
functions coincide with the Dirac observables of the continuum theory,
$O^C$. In order to see this, we note that
\begin{gather} \label{observables}
\{O^D,H\}=\sum_k \phi_k \{O^D,\phi_k\}=0.
\end{gather}
Let us take a point in phase space close to the constraint surface
such that  $\phi_1 = \delta$ and $\phi_k=0$ $\forall\, k \neq 1$. From
(\ref{observables}) we therefore see that  $\{O^D,\phi_1\}=0$ at that
point. Taking $\delta \to 0$ we conclude that  $\{O^D,\phi_1\}=0$ for
points on the constraint surface. Repeating this for the other
constraints completes the proof. Conversely, to every Dirac observable
of the continuum theory corresponds one or several constants of the
motion of the discrete theory.

Let us conclude the classical part noting that in the case in which
the constraints are Abelian it is possible to give an expression for
the type 1 generating function of the canonical transformation that
gives the discrete evolution. If the
constraints $\phi_k$ are Abelian, then the discrete evolution be given
by the Hamiltonian (\ref{H}) can be understood as: ``evolve for a time
$\lambda_1$ with the f\/ield~$X_{\phi_1}$; then evolve for a time
$\lambda_2$ with the f\/ield~$X_{\phi_2}$; etc.'' Here
$\lambda_i(q,p)=\left. \partial f/\partial x_i
\right|_{x_k=\phi_k(q,p)}$ is the Lagrange multiplier of the discrete
theory. This can be done because the f\/ields commute and because
$\lambda_i(q,p)$ is constant throughout the evolution.

This construction corresponds to the evolution of the discrete Lagrangian
\begin{gather*}
L(q_n,q_{n+1},\lambda_1,\ldots,\lambda_r)=
S(q_n,q_{n+1},\lambda_1,\ldots,\lambda_r)
+g(\lambda_1,\ldots,\lambda_r),
\end{gather*}
where $\lambda_i$ are the discrete Lagrange multipliers at the time
$n$, $S$ is the Hamilton principal function of the system and $g$
satisf\/ies that: 1) $g(0)=0$ and 2) the map $\lambda_i \to
\frac{\partial g}{\partial \lambda_i}$ is the inverse of the map
$x_i \to \frac{\partial f}{\partial x_i}$. In order to see this, note
that the evolution equations of the  $q$'s yield the ``exact''
evolution with given Lagrange multipliers $\lambda_k$. The equations
for the multipliers are
\begin{gather*}
0=\frac{\partial L}{\partial \lambda_i}=\frac{\partial S}{\partial \lambda_i}+\frac{\partial g}{\partial \lambda_i} .
\end{gather*}

From the Hamilton--Jacobi equation one has that $\partial S/
\partial\lambda_i=-\phi_i$ and we recover the values of the multipliers that
we had before.

Let us now discuss the quantum theory. Given how the system was put
together, there is a~natural candidate for the evolution operator
\begin{gather*}
\hat{U}=e^{-i \hat{H}}.
\end{gather*}
The continuum limit is given by the quantum states
$|\psi\rangle $ that satisfy
\begin{gather*}
\hat{U}|\psi\rangle  = |\psi\rangle ,
\end{gather*}
which is equivalent to
\begin{gather} \label{mc}
\hat{H}|\psi\rangle  =0,
\end{gather}
which corresponds, at least classically, to the condition that the
system satisfy the constraints. Condition (\ref{mc}) with $H$ given by
(\ref{H})  is the same condition used by Thiemann in his
``master constraint'' \cite{phoenix} proposal. The latter consists
in f\/inding the solutions of (\ref{mc}) as a means to f\/inding the
physical space of the theory. In that context the operator $\hat{H}$
is not interpreted as a~Hamiltonian but as a~``master constraint''
that includes all the constraints of the problem. The idea is that
states  $|\psi\rangle $ that satisfy~(\ref{mc}) will satisfy
$\hat{\phi_k}|\psi\rangle =0$ $\forall\, k$.

The existence of solutions to (\ref{mc}) depends on if zero is in the
spectrum of  $H$. If it is not it will not be possible to f\/ind
solutions and we will say that the quantum theory does not have a~continuum limit. If zero is in the spectrum there will be a~solution
and the states that satisfy~(\ref{mc}) are the physical space of
states.

In order to def\/ine an inner product in the physical space of states
one has to see if zero belongs in the discrete or continuous part of
the spectrum of $H$. In the f\/irst case, the inner product is the same
as in the kinematical Hilbert space. In the second, the states that
satisfy (\ref{mc}) are not normalizable in the kinematical inner
product and are generalized states. It is possible to introduce an
inner product in this generalized space in the following way.
One takes the projector
$\hat{P}$ from the kinematical to the physical Hilbert space. The
internal product between two states  $|\psi_1\rangle_{\rm Ph}$ and
$|\psi_2\rangle _{\rm Ph}$ of the physical Hilbert space such that  $|\psi_i\rangle _{\rm Ph}=\hat{P}|\psi_i\rangle $
with  $|\psi_i\rangle $ states of the kinematical Hilbert space is given by
\begin{gather*}
{}_{\rm Ph} \langle \psi_1|\psi_2\rangle _{\rm Ph}=\langle \psi_1|\hat{P}|\psi_2\rangle.
\end{gather*}

Such a projector can be constructed from the evolution operator in the
following manner
\begin{gather} \label{proyector}
\hat{P}\equiv \lim_{n\to\infty} C_n \hat{U}^n
\end{gather}
with appropriate constants (c-numbers)
$C_n$ such that $\lim\limits_{n\to\infty}(C_{n+1}/C_n)=1$.
One then has that $\hat{U} \hat{P} = \hat{P}$ and therefore $\hat{U}
\hat{P} |\psi\rangle  = \hat{P}|\psi\rangle$ for all states  $|\psi>$ of the
kinematic Hilbert space.

The quantization of the resulting discrete theories have important
conceptual simplif\/ications with respect to the original continuum
theory. First of all, the theories are free of constraints. The latter
are approximately enforced by the evolution and can be made as small
as one wishes by choosing initial data. But one does not need to be
concerned with issues like constraint algebras anymore. One will be
quantizing a classical theory that, although it approximates the
theory of interest with arbitrary precision, is conceptually very
dif\/ferent.

To quantize these theories we start by picking a kinematical Hilbert
space that will be given by square integrable functions of the
conf\/iguration variables $\Psi(q)$. On this space one def\/ines operators
$\hat{Q}$ and $\hat{P}$ as usual.  To construct the operators at other
time levels one introduces a~unitary
operator $\hat{U}$ that we will def\/ine later, with the following
properties
\begin{gather*}
\hat{Q}_{n} \equiv \hat{U}^{-1} \hat{Q}_{n-1} \hat{U} =
\hat{U}^{-n} \hat{Q}_0
  \hat{U}^n , \qquad \hat{P}_{n} \equiv \hat{U}^{-1}
\hat{P}_{n-1} \hat{U} =
\hat{U}^{-n} \hat{P}_0  \hat{U}^n .
\end{gather*}

The evolution operator is given by $\hat{U}=e^{-i\hat{H}/\hbar}$. The
operator $\hat{U}$ may also
be determined by requiring that the fundamental operators satisfy an
operatorial version of the evolution equations{\samepage
\begin{gather*}
\hat{Q}_{n}  \hat{U} - \hat{U} Q_{n+1}(\hat{Q}_n,\hat{P}_n)=0,
\qquad \hat{P}_n
\hat{U} - \hat{U}
P_{n+1}(\hat{Q}_n,\hat{P}_n)=0,
\end{gather*}
and this provides a consistency condition that aids in the
construction of $\hat{U}$.}

We recall that classically $H=0$ if and only if the
constraints $\phi_i=0$.  One can def\/ine naturally the
physical space of the continuum theory in a way that does not require that we
refer to the constraints directly. Since we know that $\hat{U}=\exp(-i
\hat{H}/\hbar)$, a necessary condition satisf\/ied by the states of the
physical space of the continuum theory, $\psi\in{\cal H}_{\rm phys}$
is given by $\hat{U}\psi = \psi$. More precisely the states $\psi$ of
${\cal H}_{\rm phys}$ should belong to the dual of a space $\Phi$ of
functions suf\/f\/iciently regular on ${\cal H}_k$.  That is, the states
$\psi\in {\cal H}_{\rm phys}$ satisfy
\begin{gather*}
\int \psi^* \hat{U}^\dagger \varphi dq = \int \psi^* \varphi dq,
\end{gather*}
where $\varphi\in \Phi$.
This condition characterizes the quantum physical space
of a constrained continuum theory without needing to implement the
constraints as quantum operators. This is an important advantage
in situations where implementing the constraints as quantum operators
could be problematic, as for instance when they do not form a Lie
algebra. This is the case in general relativity where the constraint
algebra has structure functions. For more details on the quantization
see~\cite{uniform}.

\subsubsection{Examples}

We discuss now two signif\/icant examples that will illustrate the main
two possibilities when implementing the construction outlined here. In
the f\/irst example, one f\/inds that the quantum master constraint
includes the zero eigenvalue in its kernel. There the technique
reproduces the results of the ordinary Dirac quantization
technique. In the second example the zero eigenvalue is not in the
kernel. The resulting quantum theory has a fundamental level of
discreteness and will only recover semiclassically the continuum
theory in certain circumstances. In this example the Dirac
quantization is also problematic. We believe this example illustrates
the most likely situation one will face in gravity: the resulting
quantum theory will have a fundamental level of discreteness and
classical general relativity will only be recovered in long length
scales compared to the Planck length. This is a point of view that
many people have advocated for an ultimate theory of quantum gravity
over the years.

The two examples are taken from~\cite{uniform}.

\subsubsection{Model with continuum limit}
Here we will consider a model constrained system with non-Abelian
constraints that form a Lie algebra. We consider a mechanical system
with conf\/iguration manifold $R^3$, coordinatized by~$q^i$, $i=1,2,3$ and
3 non-commuting ``angular momentum'' constraints
\begin{gather*}
C^i=L^i\equiv \epsilon^{ijk} q^j p^k, \qquad i=1,2,3,
\end{gather*}
where the $p^i$'s are the momenta conjugate to the $q^i$'s and we
assume Einstein's summation convention on repeated indices.  The three
constraints are not independent. Vanishing angular momentum is
equivalent to requiring $q^i=\lambda p^i$ with $\lambda$ arbitrary,
which implies two conditions to be satisf\/ied by the phase space
variables. The constraint surface is four dimensional and therefore
the system has two independent observables. We can
choose to construct the Hamiltonian starting from two
independent constraints or one could choose to use a more symmetric,
yet redundant, form of the constraints, which will be our choice.  The
Hamiltonian that will dictate evolution is
\begin{gather*}
  H=\sum_{i=1}^3 {(L^i)^2\over 2 k},
\end{gather*}
where $k$ is a constant with units of action. The resulting discrete
evolution equations are
\begin{gather*}
  q^i_{n+1}  =  q^i_n +\epsilon_{ijk}{ q^k L^j\over k}
+\epsilon_{ijk}\epsilon_{jmn} q^m {L^n L^k \over k^2} +\cdots,\\
  p^i_{n+1}  =  p^i_n +\epsilon_{ijk} {p^k L^j \over k}
+\epsilon_{ijk}\epsilon_{jmn} p^m {L^n L^k \over k^2}+\cdots.
\end{gather*}

One obtains the continuum limit by setting $H=\delta^2/2$ and def\/ining
$\lambda^i = L^i/(\delta\sqrt{k})$. Taking the limit as
explained before, one gets
\begin{gather*}
  \dot{q}^i  =  \epsilon_{ijk} {q^k \lambda^j \over \sqrt{k}},\qquad
  \dot{p}^i  =  \epsilon_{ijk} {p^k \lambda^j \over \sqrt{k}}.
\end{gather*}

The components $L^i$ are constants of the motion and therefore
$\lambda^i$ are constant. There exist three constants more: $q\cdot
q=q^iq^i$, $p\cdot p=p^ip^i$, $q\cdot p = q^ip^i$. These are not
independent since $L^2-((q\cdot q)(p\cdot p)-(q\cdot p)^2)=0$.  One
therefore has f\/ive independent constants of the motion of the discrete
theory.  In the continuum limit the $L^i$'s vanish and one has two
independent constants of motion, for instance $q\cdot q$ and $q\cdot
p$.  In the continuum, the trajectories are arbitrary trajectories on
a sphere. In the discrete theory, when one takes the continuum limit
one obtains trajectories that correspond to arbitrary circumferences
on the sphere, since the $\lambda^i$'s are constant. The constraint
surface is therefore completely covered, but not all orbits of the
continuum theory are recovered. This however, is not problematic since
we can recover all physical information with the trajectories obtained.

To quantize the system, we will start with an auxiliary Hilbert space
$L_2(R^N)$ on which the operators $\hat q^i$ act as multiplication
operators and the momenta as derivatives
\begin{gather*}
\hat q^i  \psi(q) = q^i
 \psi(q),\qquad
\hat p^i\psi(q) = -i\hbar  \partial_i \psi(q),
\end{gather*}
and immediately construct the unitary operator
\begin{gather*}
  \hat{U}=\exp\left(-i {\hat{L}^2 \over 2 k \hbar}\right),
\end{gather*}
which recovers (up to terms of order $\hbar$) the classical discrete
evolution equations as operatorial relations so the correspondence
principle is satisf\/ied.

To compute the projector we use the formula (\ref{proyector}) and,
work with a basis labeled by  the radial and angular momentum
eigenvalues. The projector is given by
\begin{gather*}
  \hat{P} |n,\ell,m\rangle  = \delta_{\ell 0} |n,\ell,m\rangle  =\delta_{\ell 0} |n,0,0\rangle ,
\end{gather*}
which can be rewritten as,
\begin{gather*}
  \hat{P} = \sum_{n=0}^\infty |n,0,0\rangle \langle 0,0,n|.
\end{gather*}

The continuum limit is therefore immediately achieved. The physical
space is the space of vanishing angular momentum and on it the constraints
of the continuum theory are automatically satisf\/ied. We have therefore
recovered the usual Dirac quantization.  The physical space is
given by the square integrable functions depending on the radial
variable $|q|$. The physical inner product is therefore
induced by the kinematical inner product on the physical space of
states.

\subsubsection{A model with non-compact gauge group}

We will now consider a system with non-Abelian constraints that form a
Lie algebra that is non-compact, associated with the $SO(2,1)$
group. It has the same phase space as in the previous example, except
that the metric will be given by ${\rm diag}(-1,1,1)$.
The constraints are
\begin{gather*}
C_i=L_i\equiv \epsilon_{ij}{}^k q^j p_k, \qquad i=0,1,2.
\end{gather*}
One has to be careful about upper and lower indices since the
metric is non-trivial. Just like before,
the three constraints are not independent and the constraint surface
is four dimensional and the system has two independent
observables. The
canonical evolution is given through the master constraint
\begin{gather*}
  H= {L_i L^i+2 L_0^2\over 2 k},
\end{gather*}
where $k$ is a constant with units of action. The discrete
evolution equations are
\begin{gather*}
  q^i_{n+1}  =  q^i_n +\epsilon^i{}_{jk}
{q^k L^j\over k}-2\epsilon^i{}_{0k} {q^k L^0\over k} +\cdots,\\
  p^i_{n+1}  =  p^i_n +\epsilon^i{}_{jk}
{p^k L^j\over k}-2\epsilon^i{}_{0k} {p^k L^0\over k}+\cdots.
\end{gather*}

The continuum limit is achieved setting $H=\delta^2/2$. Def\/ining
$\lambda^i = L^i/(\delta\sqrt{k})$ for $i=1,2$ and $\lambda^0
=-L^0/(\delta\sqrt{k})$ we have, taking the limit as explained
before
\begin{gather*}
  \dot{q}^i  =  \epsilon^i{}_{jk} {q^k \lambda^j \over \sqrt{k}},\qquad
  \dot{p}^i  =  \epsilon^i{}_{jk} {p^k \lambda^j \over \sqrt{k}}.
\end{gather*}

We note that although $L_0$ is a constant of the motion, the other
components are not. However $L_1^2+L_2^2$ is a constant, so $L_1$
and $L_2$ rotate around $L_0$ throughout the evolution.  There
exist three further constants $q\cdot q=q^i q_i$, $p\cdot
p=p^ip_i$ $q\cdot p = q^ip_i$. These are not independent since
$L\cdot L-((q\cdot q)(p\cdot p)-(q\cdot p)^2)=0$.  One therefore has
four independent constants of the motion of the discrete theory.
In the continuum limit two  vanish and one is left with
two constants of motion, for instance $q\cdot q$ and $q\cdot p$.
The trajectories in the continuum are arbitrary
trajectories on two hyperboloids, one space-like and one
time-like. The continuum limit of the discrete theory yields
trajectories that correspond to particular
choices of the Lagrange multipliers, depending on the initial
conditions chosen.

The quantization of this model (and similar ones) is known to
have subtleties \cite{phoenix, gomberoffmarolf, loukorovelli}. The
central problem is that if one promotes the constraints to operators on
the usual Hilbert space, they do not have a vanishing eigenvalue in
their spectra. This
can happen, but usually the resolution is to extend the Hilbert space
by including an improper basis of eigenvectors. This can be done in
this case, but one f\/inds that the spectrum again does not contain
zero. More specif\/ically, the continuum spectrum has eigenvalues larger
than $\hbar^2/4$.  One can f\/ind eigenvectors with eigenvalues smaller
than $\hbar^2/4$, but they do not arise as limit of functions of the
Hilbert space, i.e.  they cannot form part of an improper basis of the
Hilbert space. One can adopt the point of view that nevertheless the
eigenvectors with zero eigenvalue are the ``physical space'' of the
theory, essentially abandoning the idea that the physical space arises
as a suitable limit of the kinematical space. This was suggested in~\cite{gomberoffmarolf,loukorovelli}.  From the point of view of our
approach this does not suf\/f\/ice, since we wish to construct the
physical space of states starting from the quantum kinematical space,
taking a limit. As we mentioned before, lacking the zero eigenvalue in
the spectrum of the Hamiltonian yields our prescription for the
projector on the physical space useless.  With this in mind, the most
satisfactory solution is the one chosen in~\cite{phoenix}, where one
chooses as physical space the eigenvectors that have $\hbar^2/4$ as
eigenvalue. Another attractive possibility in the discrete approach is
{\em not} to take the continuum limit and retain a~level of
fundamental discreteness. This is very natural in a model where it is
dif\/f\/icult to achieve a~vanishing eigenvalue for the constraints, a
natural minimum existing for their eigenvalues. As we argued before,
such models can approximate the semiclassical physics of the theory of
interest with some restrictions on the type of states considered.
There can therefore be viewed as the best thing one can do in terms of
having an underlying quantum theory for the model that approximates
the classical physics of interest.

\subsubsection{Other examples}

The above two examples capture the essence of the technique in which
we see in one case that we recover the traditional Dirac quantization,
whereas in the other case, where even the Dirac procedure faces
issues, we obtain a satisfactory solution. The technique has also been
applied in $2+1$ gravity \cite{uniform} where one recovers the traditional
quantization and more recently to the study of spherically symmetric
gravity coupled to a spherically symmetric scalar f\/ield
\cite{spherical}. This is a challenging example with a rich and
complex dynamical structure and inf\/inite degrees of freedom. Up to now
only low energy regimes close to f\/lat space have been studied . One
cannot achieve a continuum limit and one is left with a theory with
a fundamental level of discreteness that nevertheless approaches
semiclassically general relativity very well, as we had anticipated.

\section{Improved and perfect actions}\label{Sec:ImprovedAndPerfectActions}

In this section we review the mechanism of symmetry breaking in the covariant language, i.e.~within Regge Calculus. We also describe the perfect action approach in order to cope with the arising problems, and demonstrate the connections to the renormalization group.

\subsection{Dif\/feo breaking in Regge}

\begin{figure}[t]
\centering
\includegraphics{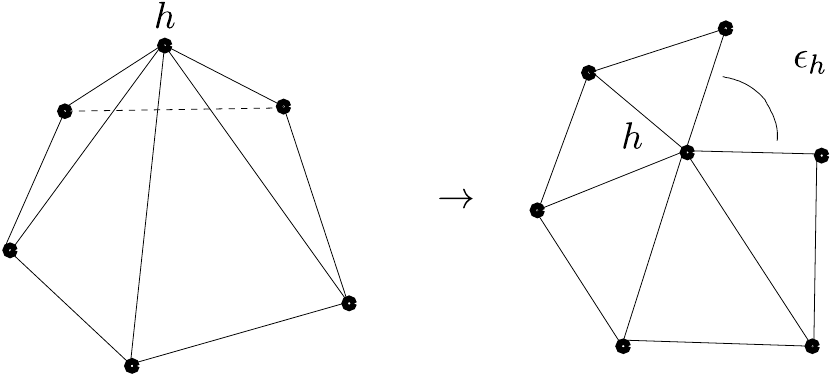}
\caption{Def\/icit angle $\epsilon_h$ situated at a $(d-2)$-dimensional simplex (hinge) $h$.}\label{Fig:Figure01}
\end{figure}

We consider the discretization of General Relativity via Regge's method, and demonstrate in which sense the dif\/feomorphisms of the continuum are broken. In Regge calculus $d$-dimensional space-time is discretized by a triangulation $\mathcal{T}$ consisting of simplices $\sigma$, and the metric information is encoded as edge lengths\footnote{For simplicity we consider only Riemannian metrics in what follows, although Lorientzian metrics can equally well be treated by considering the edge length squares $l_e^2$, which can e.g.~be negative if the edge is time-like. Special care has to be taken when considering the causal structure of the triangulation~\cite{NULLSTRUT}.} $l_e$ of the simplices $\sigma$. By specifying the edge lengths the geometry of the simplex is uniquely determined, and the discrete analogue of the Einstein--Hilbert action~(\ref{Gl:EHAction}), called the Regge action, is given by
\begin{gather}\label{Gl:ReggeAction}
S_{\text{R}}[l_e] = \sum_{h\in \mathcal{T}^\circ} V_h\underbrace{\left(2\pi-\sum_{\sigma\supset h}\theta_h^\sigma\right)}_{=:\epsilon_h} + \sum_{h\in \partial\mathcal{T}} V_h\underbrace{\left(\pi-\sum_{\sigma\supset h}\theta_h^\sigma\right)}_{=:\psi_h} - \Lambda\sum_\sigma V_\sigma ,
\end{gather}
where the f\/irst sum ranged over all $d-2$-simplices (called hinges) in the interior, and the second sum ranges over all hinges in the boundary of $\mathcal{T}$. The curvature is encoded in the excess of the sum of dihedral angles  $\theta_h^\sigma$ around a hinge~$h$ with respect to $2\pi$ (see Fig.~\ref{Fig:Figure01}), and denoted as def\/icit angle~$\epsilon_h$.
Similarly, for a hinge $h\in\partial\mathcal{T}$ on the boundary, $\psi_h$ encodes the extrinsic curvature of the boundary surface $\partial\mathcal{T}$ in $\mathcal{T}$. Also, $V_h$ denotes the $(d-2)$-dimensional volume of the hinge $h$, and~$V_\sigma$ the $d$-dimensional volume of the simplex $\sigma$.

The boundary terms and the Schlaef\/li-identity \cite{SCHLENKER} ensure that the equations of motion are given by
\begin{gather}\label{Gl:ReggeEquations}
\frac{\partial S_{\text{R}}}{\partial l_e}\;=\;\sum_{h\supset e}\frac{\partial V_h}{\partial l_e}\epsilon_h -
\Lambda\sum_{\sigma\supset e}\frac{\partial V_\sigma}{\partial l_e} = 0
\end{gather}
 for all interior edges $e\in\mathcal{T}^\circ$, which are conditions for the interior edge lengths $l_e$ with f\/ixed boundary edge lengths.

In the continuum theory, i.e.~GR, the Einstein equations do not determine uniquely a solution for the interior metric, but only up to dif\/feomorphism. Also, the Einstein equations contain as subset consistency conditions for the boundary metric, which do not contain dynamical information, but constitute constraints. In the discrete setting the same situation appears in the special case of $d=3$ and $\Lambda=0$: The hinges $h$ and the edges~$e$ coincide, and therefore the Regge equations (\ref{Gl:ReggeEquations}) simply become
\begin{gather}\label{Gl:ReggeEquations3DL=0}
\epsilon_e = 0.
\end{gather}
For given boundary lengths $l_e$, the equations (\ref{Gl:ReggeEquations3DL=0}) have in general inf\/initely many solutions for the interior lengths, since there are inf\/initely many ways of f\/illing up a portion of f\/lat $3$-dimensional space with tetrahedra. Of a given solution for the $l_e$, another solution can be constructed by vertex translation, and it can be shown that this transformation between solutions is generated by the discrete Bianchi-identities \cite{HERBIEBIANCHI,Dittrich01, Freidel:2002dw}, which therefore corresponds exactly to a dif\/feomorphism in this case. Also, as part of the Regge equations, the boundary lengths have to satisfy a constraint, which demands that the $2d$ boundary triangulation can be immersed in f\/lat $3d$ space~\cite{Dittrich:2007wm}, which is again completely similar to the continuum case.

However, as soon as one leaves the case of $d=3$ or $\Lambda=0$, the situation changes. This can be seen as follows: The non-uniqueness of the solutions to (\ref{Gl:ReggeEquations3DL=0}) manifest themselves in zero Eigenvalues of the Hessian matrix $H_{ee'}=\partial^2S_{\text{R}}/\partial l_e\partial l_{e'}$. These zeros are precisely what makes the Legendre transformation singular, indicating the presence of gauge symmetries \cite{bahrdittrich1}. The same zeros appear in the Hessian for~$4d$, $\Lambda=0$, whenever evaluated on a f\/lat solution $\epsilon_t=0$ (where $t$ denotes the triangles of~$\mathcal{T}$), and the zero modes correspond to the dif\/feomorphism gauge modes of linearized gravity~\cite{Rocek}.

\begin{figure}[t]
\centering
\includegraphics{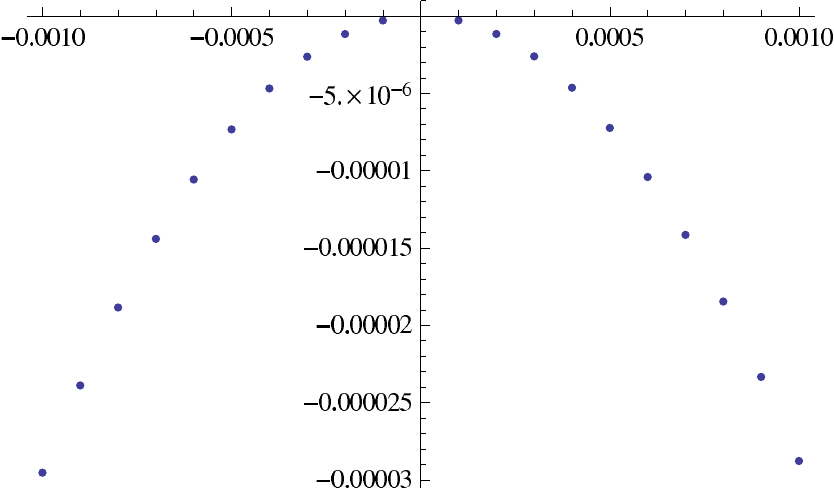}
\caption{The lowest eigenvalue of the Hessian as a function of deviation from f\/latness (proportional to~$\epsilon$), in a particular $4d$-triangulation with one inner vertex, discussed in \cite{bahrdittrich1}.} \label{Fig:Hessian5v}
\end{figure}

It was well-known for some time that for $4d$ and $\Lambda=0$ the vertex displacement symmetry, around the f\/lat solution, is satisf\/ied at least up to terms of order $\epsilon^2$ \cite{Hamber:1992df,Rocek}. In \cite{bahrdittrich1} this bound was made sharp by showing that the Hessian is non-degenerate, i.e.~has no zero Eigenvalues, in the case of~$3d$, $\Lambda\neq0$, and in~$4d$, $\Lambda=0$ whenever the boundary data is such that the solution has internal curvature. In fact, the smallest Eigenvalue scales with~$\epsilon^2\sim\Lambda^2$ in the case of $3d$, as well was with $\epsilon^2$ (where $\epsilon$ is a def\/icit angle at an internal edge) in $4d$ (see Fig.~\ref{Fig:Hessian5v}). As soon as the solution to~(\ref{Gl:ReggeEquations}) exhibits curvature, there is no continuous symmetry relating dif\/ferent solutions anymore, and although in some cases there are dif\/ferent discrete solutions~\cite{Piran:1985is}, these are rather a discretization artifact. In most cases the solution for given boundary data is unique.
  Moreover, the part of the Regge equations which, in the f\/lat case, constituted the constraints, i.e.~which involved only the initial data on one time step\footnote{Here the dif\/ferent time steps were taken as a series of so-called tent moves~\cite{Barrett:1994ks}. For a canonical framework for simplicial theories which builds on Pachner moves, see also~\cite{hohn,Dittrich:2008ar,NewAngle,Dittrich:2011ke}.}, acquired a weak coupling between data on subsequent time slices, i.e.~turned from a constraint into a (weakly) dynamical equation. These equations are therefore called \emph{pseudo-constraints}.

The uniqueness of solutions in cases with nonvanishing curvature (i.e.~the breaking of the continuum gauge symmetry) leads to a vast overcounting of degrees of freedom.
Consider e.g.~the case of $3d$ Regge with $\Lambda\neq 0$. For a f\/ixed triangulation, since there are no constraint equations for the boundary data, not only can one prescribe many more initial conditions, each of which lead to a unique solution of motion, since there are inf\/initely many dif\/ferent triangulations (even with the same boundary triangulation), each of which produce physically dif\/ferent solutions, the discrete theory has many more than just the f\/initely many (topological) degrees of freedom that the continuum theory has.

The breaking of symmetries in the discrete therefore leads to the emergence of pseudo-degrees of freedom, leading to several problems: The f\/irst is an interpretational one, which has also been realized in other discrete, in particular numerical, approaches to GR \cite{Lehner:2001wq}: It is unclear how to distinguish the pseudo degrees of freedom from the actual physical ones. It is therefore dif\/f\/icult to extract the physical content from a numerical solution, e.g.~disentangling actual oscillations of gravitational waves from oscillations of the chosen coordinate system.

\looseness=-1
Another problem arises within any attempt to build a quantum gravity theory based on Regge discretizations: Generically, breaking of gauge symmetries within the path integral measure leads to anomalies, in particular problematic in interacting theories \cite{'tHooft:1972fi}. In quantum theories based on Regge calculus, the path integral will, for a very f\/ine triangulation, contain contributions of lots of \emph{almost} gauge equivalent discrete metrics, each with nearly the same amplitude. Hence the amplitude will not only contain inf\/inities of the usual f\/ield theoretic nature, but also coming from the ef\/fective integration over the dif\/feomorphism gauge orbit. This is in particular a problem for the vertex expansion of the spin foam path integral, as advocated in~\cite{Henderson:2010qd}. The triangulations with many vertices will contribute much more to the sum than the triangulations with few vertices, so that, at every order, the correction terms dominate, rendering the whole sum severely divergent. We will see this phenomenon occur in Section~\ref{Sec:CoarseGrainingQuantum}, in the case of a toy model.

In the following we describe an attempt to circumvent the above problems, by constructing a discrete action with the correct amount of symmetries. The main technical tool for this is coarse graining.

\subsection{The coarse graining idea: classical}

A toy model for the breaking of dif\/feomorphism symmetry in Regge calculus, which has been investigated in \cite{Bahr:2009qc,Bahr:2011uj,Rovelli:2011fk} is parameterized mechanics, in which time $t$ itself is treated as a~conf\/iguration variable, which, together with $q$, evolves with respect to an auxiliary parameter~$s$. The broken symmetry in the discrete case is invariance under reparametrization $s\to s'(s)$, which is ef\/fectively dif\/feomorphism symmetry in $1d$. In this case, there is a way of constructing an action for the discrete theory (i.e.~depending on $(t_n,q_n)$), which nevertheless exhibits the exact symmetry from the continuum theory. Given a discretization of the continuum action~$S^{(0)}$, this \emph{perfect action} can be constructed iteratively by a coarse graining procedure, resulting in a~sequence of discrete actions~$S^{(n)}$ converging to the perfect action $S^{(\infty)}$ in the limit.

\begin{figure}[t]
\centering
\includegraphics{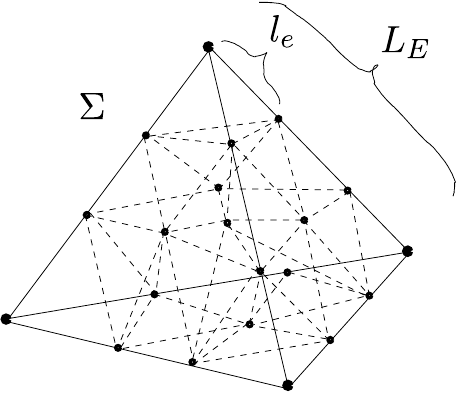}
\caption{A tetrahedron $\Sigma$ subdivided into smaller tetrahedra~$\sigma$. The length~$L_E$ of a coarse edge~$E$ is the sum of all the lengths~$l_e$ of edges $e\subset E$ that constitute~$E$.}\label{Fig:CoarseGrained}
\end{figure}

A similar construction can be employed in principle within the context of Regge calculus, and we show how this procedure, described in~\cite{Bahr:2009qc}, is able to cure the problems described above in the case of $3d$ Regge with $\Lambda\neq 0$.
Consider a $3d$ triangulation $\mathcal{T}$ (capturing the macroscopic degrees of freedom) and a ref\/ined triangulation $\tau$, such that each edge $E\in\mathcal{T}$ can be composed of edges $e\in\tau$ (see Fig.~\ref{Fig:CoarseGrained}), capturing the microscopic degrees of freedom. The ef\/fective (or \emph{improved}) action $S_{\mathcal{T},\tau}[L_{E}]$, which lives on $\mathcal{T}$, is constructed by putting the Regge action on $\tau$ and integrating out the microscopic degrees of freedom. To this behalf, one solves the Regge equations for the $l_{e}$, $e\in\tau$, with f\/ixed edge lengths
\begin{gather}\label{Gl:ConstraintOnEdgeLengths3DLambda}
\sum_{e\subset E}l_{e}\;\stackrel{!}{=}\;L_E .
\end{gather}
 One therefore adds the conditions (\ref{Gl:ConstraintOnEdgeLengths3DLambda}) to the Regge action (\ref{Gl:ReggeAction}) via Lagrange multipliers $\alpha_E$, extremizing
\begin{gather}\label{Gl:ActionWithLagrangeMultipliers3DLambda}
S[l_{e}] = \sum_{e}l_{e}\epsilon_{e} - \Lambda\sum_{\sigma\in\tau}V_{\sigma} + \sum_E\alpha_E\left(L_E-\sum_{e\subset E}l_{e}\right) .
\end{gather}
Note that the action (\ref{Gl:ActionWithLagrangeMultipliers3DLambda}) depends on the $L_E$ as parameters. The ef\/fective action $S_{\mathcal{T},\mathcal{T}'}[L_E]$ is then def\/ined as the value of the action (\ref{Gl:ActionWithLagrangeMultipliers3DLambda}) evaluated on a solution, i.e.
\begin{gather*}
S_{\mathcal{T},\tau}[L_E] := S[l_{e}]_{\Big|\frac{\partial S}{\partial l_{e}}=\frac{\partial S}{\partial \alpha_E}=0} .
\end{gather*}
 One can show straightforwardly that
\begin{gather*}
S_{\mathcal{T},\tau}[L_E] = \sum_E L_E\alpha_E + 2\Lambda\sum_{\Sigma\in\mathcal{T}}V_\Sigma
\end{gather*}
 with $V_\Sigma:=\sum\limits_{\sigma\subset\Sigma}V_{\sigma}$ for a simplex $\Sigma\in\mathcal{T}$. Note that both $\alpha_E$ and $V_\Sigma$ are a priori complicated functions of the $L_E$, which have to be determined by the equations of motion for $S$. Without knowing their form explicitly, setting
\begin{gather*}
\Theta_E^\Sigma := \sum_{\sigma\supset e,\sigma\subset\Sigma}\left(\theta_e^\sigma - \Lambda\frac{\partial V_\sigma}{\partial l_e}\right),
\end{gather*}
which does not depend on the choice of $e\subset E$ by the equations of motion, one can derive the identities
\begin{gather}\label{Gl:Alpha}
\alpha_E = 2\pi - \sum_{\Sigma\supset E}\Theta_E^\Sigma, \\
\label{Gl:SchlaefliIdentityForImprovedQuantities}
\sum_{E\subset \Sigma}L_E\frac{\partial \Theta_{E}^\Sigma}{\partial L_{E'}} = 2\Lambda\frac{\partial V_\Sigma}{\partial L_{E'}} .
\end{gather}
Note that (\ref{Gl:SchlaefliIdentityForImprovedQuantities}) implies that the equations of motion for the $L_E$ as given by the ef\/fective action~$S_{\mathcal{T},\tau}$ are simply
\begin{gather*}
\frac{\partial S_{\mathcal{T},\tau}}{\partial L_E} = \alpha_E = 0 .
\end{gather*}
 Not only is (\ref{Gl:SchlaefliIdentityForImprovedQuantities}) precisely of the form of the Schlaef\/li identity for simplices of constant curvature~$\Lambda$~\cite{SCHLENKER}, but one can also show \cite{Bahr:2009qc} that in the limit of $\tau$ becoming very f\/ine with respect to $\mathcal{T}$ in a controlled way, the functions $\Theta_E^\Sigma$, $V_\Sigma$ converge to the dihedral angle and the volume within a~tetrahedron of constant curvature $\Lambda$, with edge lengths $L_E$. So the \emph{perfect action}, as the limit of inf\/inite ref\/inement~$\tau\to\infty$, is given by the Regge action with simplices of constant curvature~$\Lambda$
\begin{gather}\label{Gl:ReggeActionCurvedSimplices3D}
S_{\mathcal{T},\infty} = \sum_{E}L_E\epsilon_E + 2\Lambda\sum_{\Sigma}V_\Sigma .
\end{gather}
 Indeed, the equations of motion for (\ref{Gl:ReggeActionCurvedSimplices3D}) are simply $\epsilon_E=0$, so the solutions describe a metric of constant curvature everywhere. Just as in the case for $\Lambda=0$, where $\epsilon_E=0$ allows for arbitrary subdivision of f\/lat space into f\/lat simplices, resulting in the vertex displacement symmetry, in the $\Lambda\neq 0$ case $\epsilon_E=0$ allow for arbitrary subdivision of constantly curved space into constantly curved simplices, and the same vertex displacement symmetry appears. Hence, unlike the Regge action~(\ref{Gl:ReggeAction}), the perfect action (\ref{Gl:ReggeActionCurvedSimplices3D}) exhibits the correct symmetries of the continuum theory\footnote{In fact, (\ref{Gl:ReggeAction}) for $3d$ consists of the f\/irst two terms in the $\Lambda$-expansion of (\ref{Gl:ReggeActionCurvedSimplices3D}), i.e.~up to $O(\Lambda^2)$.}.

The same coarse graining procedure works for the $\Lambda\neq 0$ sector in $4d$, which is such that the solution to the equations of motion results in constant curvature~\cite{Bahr:2009qc}: also in this case the Regge action~(\ref{Gl:ReggeAction}) converges to its analogue with constantly curved $4$-simplices by means of the procedure described above.

\subsection{Feasibility}

The procedure for constructing the perfect action for a discrete system requires the solution of the equations of motion for a very f\/ine discretization. Even for the case for $3d$, $\Lambda\neq 0$ described above there is no analytic formula for the solutions. The simplicity of the system, and in particular the structural equation (\ref{Gl:SchlaefliIdentityForImprovedQuantities}) allowed to prove that the sequence of ef\/fective actions $S_{\mathcal{T},\tau}$ converges to (\ref{Gl:ReggeActionCurvedSimplices3D}) in the limit of inf\/initely f\/ine $\tau$. For more complicated systems (in particular for discrete GR in $4d$), it might not be feasible to try and compute the perfect action explicitly. In particular, in the known cases the perfect actions are always closely related to the Hamilton's principal function of the continuous system \cite{Bahr:2009qc}, the explicit knowledge of which is equivalent to having control over the solution space of the continuum theory, which is def\/initely not the case for $4d$ GR.

Still, the concept of a perfect action is a very powerful tool in numerical studies of lattice QCD \cite{Hasenfratz:1997ft,Bietenholz:1999kr}, where it is used to minimize discretization artifacts. Here, the perfect action is not known explicitly, but there exist very good numerical and analytical approximations to it. In \cite{Bahr:2010cq}, a procedure was described for how to compute the perfect action approximately within a vicinity of a point in conf\/iguration space. Here it was found that, if one concentrates on the perfect action around a solution $\{l_e^{(0)}\}$ which exhibits the exact symmetry of the continuum (as e.g.~the case for f\/lat space in $4d$ Regge for $\Lambda=0$), the gauge symmetries are satisf\/ied to linear order in the perturbative expansion. However, the breaking of the gauge symmetries to higher order imposes consistency conditions on the actual state around which one perturbs. Within the gauge equivalence class $\{l_e^{(0)}\}$ of solutions which all solve the equations of motion, there is only one particular solution around which the perturbative expansion is consistent. In fact, this solution is singled out by the condition that the Hamilton's principal function to second order depends minimally on the gauge parameters \cite{hohn}.

Computing the perfect action to lowest non-trivial order for linear systems has been performed in \cite{Bahr:2010cq}, where it could be shown that, for linearized gravity the perfect action (to linear order) is invariant under coarse graining, and displays the correct amount of symmetries in~$3d$. In~$4d$, there appear non-localities, which make the results more complicated, and which could in principle be dealt with by Migdal--Kadanof\/f-type approximation schemes \cite{Migdal:1975zg,Dittrich:2011zh}.

\subsection{Coarse graining: quantum}\label{Sec:CoarseGrainingQuantum}

The broken dif\/feomorphism symmetry of discretized theories can lead to interpretational problems in the classical realm, as we have argued in the last section. Here one cannot easily distinguish the physical degrees of freedom from the ones that arise by breaking the gauge symmetry. The f\/iner the discretization, the better the approximate symmetries (i.e.~the closer the minimal Eigenvalues in the Hessian are to actually vanish), and the closer the pseudoconstraints are to the actual constraints of the continuum theory.

Within the corresponding quantum theory, the problem becomes even more severe. For f\/inite discretization, the additional degrees of freedom, which become gauge in the continuum limit, are being treated as physical and are summed over in the path integral, which lead to unphysical singularities\footnote{As opposed to e.g.~UV or IR divergencies in QFT, which carry information about the physical properties of the system.}.

Consider for instance the example of parametrized $1d$ mechanics, with continuum variables~$q(s)$ and~$t(s)$, $s$ being an auxiliary gauge parameter. Due to the gauge symmetry (i.e.~re\-pa\-ra\-met\-ri\-za\-tion-invariance) only the dependence $q(t)$ is physical, rather than $q$ and $t$ itself, and one indeed gets the correct answer for the propagator if one drops the integration over $t$ within the path integral, and only integrates over $q$.

The discrete conf\/iguration variables are $(t_n, q_n)$ and $n=1,\ldots, N$. In the path integral
\begin{gather}\label{Gl:NaiveDiscretePropagator}
K^{(0)}(t_0,q_0,t_n,q_n) = \int_{\mathbb{R}^{2N}} dt_n dq_n\;e^{\frac{i}{\hbar}S_{\text{discrete}}(t_n,q_n)}
\end{gather}
the summation over both $t_n$ and $q_n$ have to be performed, since there is no gauge symmetry between the variables, if one uses a generic discretized approximation to the continuum action. In the limit of f\/iner and f\/iner discretization (i.e.~letting $N$ go to inf\/inity), the contribution coming from the integration over the $t_n$ approaches the integral over the gauge group (which is inf\/inite\footnote{This system has also bee investigated in \cite{Rovelli:2011fk}, where the monotonicity of the $t_n$ was enforced, i.e.~it was only integrated over $t_0<\cdots <t_k<t_{k+1}<\cdots <t_N$, which improves convergence properties. In~\cite{Bahr:2011uj} the integration range for each $t_k$ was the whole real line however, since this resembles more closely what happens in e.g.~the Ponzano--Regge model~\cite{PROR, PR} where a similar restriction would be very dif\/f\/icult to implement.}). While in this particular case the problem can be solved by just neglecting the $t_n$-integration by hand, this relies on an apriori knowledge of which variables remain physical and which become gauge within the continuum, i.e.~a~separation of the conf\/iguration variables into~$t$ and~$q$. For more complicated systems (in particular within GR), this separation is not possible.

However, given the propagator $K^{(0)}$ as by (\ref{Gl:NaiveDiscretePropagator}), it is possible to construct a series of propagators $K^{(n)}$, which converge to a propagator $K^{(\infty)}$ which avoids the problems above and which is the quantum analogue of the perfect action. This is done by performing the quantum analogue of the coarse graining procedure described in the last section, and closely resembles the construction of a Wilsonian renormalization group f\/low:

For a given discretization $2N$ one sets up the propagator (\ref{Gl:NaiveDiscretePropagator}), and performs the integration over every odd $t_{2k+1}$, $q_{2k+1}$, keeping the even $t_{2k}$, $q_{2k}$ f\/ixed. For $N=1$ this simply reads
\begin{gather}\label{Gl:WilsonianRG1DMechanics}
K^{(n+1)}(t_0,q_0,t_2,q_2) := \int_{\mathbb{R}^2}dt_1,dq_1 K^{(n)}(t_0,q_0,t_1,q_1) K^{(n)}(t_1,q_1,t_2,q_2) .
\end{gather}
This procedure has been tested for the harmonic oscillator\footnote{Note that due to the discretization, this system is not trivial, since the action is non-polynomial in the $t_n$.} and for the quartic anharmonic oscillator (to linear order in the interaction parameter~$\lambda$) in~\cite{Bahr:2010cq}. It could be shown that in the limit~$n\to\infty$ the~$K^{(n)}$ converge to an inf\/inite contribution coming from the $t_n$ integration, times a propagator $K^{(\infty)}$.

The perfect propagator $K^{(\infty)}$ satisf\/ies some crucial properties:
\begin{itemize}\itemsep=0pt
\item $K^{(\infty)}$ exhibits the correct gauge symmetry, i.e. it is invariant under a simultaneous transformation of e.g.~$t_0$ and $q_0$ via
    \begin{gather}\label{Gl:SymmetryOfPerfectPropagator}
    t_0 \to \tilde t_0 = t_0 + \Delta t_0, \qquad
    q_0 \to \tilde q_0 = q_0 + \Delta q(t_0,q_0,\Delta t_0),
    \end{gather}
 which are such that $(\tilde t_0, \tilde q_0)$ lie on a solution to the continuum equations of motion connecting $(t_0,q_0)$ and $(t_1,q_1)$.
\item Equivalently, $K^{(\infty)}$ satisf\/ies the correct constraint equation, i.e.\
    \begin{gather*}
    \left(i\hbar\frac{\partial}{\partial t_0} -  H \left(q_0,\frac{\partial}{\partial q_0}\right)\right)K^{(\infty)}(t_0,q_0,t_1,q_1) = 0,
    \end{gather*}
  where $H(q,p)$ is the Hamiltonian of the system.
\item $K^{(\infty)}$ is a f\/ixed point of the recursion relation (\ref{Gl:WilsonianRG1DMechanics}), if one neglects the $t_1$-integration. Due to the symmetry (\ref{Gl:SymmetryOfPerfectPropagator}) it is clear that the integral over $t_1$ leads to a divergent result, and this is exactly the volume of the orbit of the discrete gauge symmetry.
\item Parametrizing the propagator by $K^{(n)} = \eta^{(n)} e^{\frac{i}{\hbar}S^{(n)}}$, i.e.~an action and a measure factor, the relation (\ref{Gl:WilsonianRG1DMechanics}) leads to renormalization group equations for~$\eta^{(n)}$ and~$S^{(n)}$. One can show that in each case at the f\/ixed point~$S^{(\infty)}$ agrees with the perfect action for the system, and~$\eta^{(\infty)}$ can be regarded as the perfect path integral measure, which contains the quantum corrections to the perfect action, leading to the correct symmetries of $K^{(\infty)}$ ref\/lected in~(\ref{Gl:SymmetryOfPerfectPropagator}).
\item The quantum theory def\/ined by $K^{(\infty)}$ is discretization-independent, in the sense that it does not depend on $N$, i.e.~performing the same calculation for any other $N$ leads back to the same propagator $K^{(\infty)}$.
\end{itemize}

  It is not hard to see that the last property is a necessary result of the invariance under gauge symmetries (\ref{Gl:SymmetryOfPerfectPropagator}): Assume there is a propagator $K_N(t_i,q_i,t_f,q_f)$, which for $N=1$ time steps satisf\/ies an invariance property such as (\ref{Gl:SymmetryOfPerfectPropagator}) for both $(t_i,q_i)$ and $(t_f,q_f)$. Then the propagator for $N=2$ time steps is given by
\begin{gather}\label{Gl:ProofOfDiscretizationIndependence1D}
K_2(t_i,q_i,t_f,q_f) = \int_\mathbb{R}dq\,K_1(t_i,q_i,t,q) K_1(t,q,t_f,q_f),
\end{gather}
where the integration over $t$ has been dropped since it is ef\/fectively the integration over the gauge orbit. By the property~(\ref{Gl:SymmetryOfPerfectPropagator}) for $K_1$, the right hand side of~(\ref{Gl:ProofOfDiscretizationIndependence1D}) does not depend in~$t$. Therefore we can let it go to one boundary value, i.e.~$t\to t_i$. But if the dynamics is to be consistent, the propagator $K_N$ has to satisfy
\begin{gather*}
K_N(t,q_i,t,q_f) = \delta(q_i-q_f),
\end{gather*}
  which, together with (\ref{Gl:ProofOfDiscretizationIndependence1D}), results in $K_1\equiv K_2$. By induction one can easily show that the propagators for any number $N$ of time step $K_N$ is equal to $K_1$. In this sense, $K_N$ is discretization-independent, and one can always set $N=1$.

\subsection{Renormalization}

The analysis of systems in which the discretization breaks the continuum dif\/feomorphism symmetry suggests that renormalization methods play a crucial role in any attempt of regaining it. In particular, in the examples observed above, it is only at the RG f\/ixed point where dif\/feomorphism symmetry is restored in the discrete system, such that the correct number of degrees of freedom is quantized.

This is dif\/ferent from the situation encountered in usual lattice gauge theories, and is due to the fact that the way in which the gauge symmetry and the dynamics interact dif\/ferently in this case. In lattice gauge theories, one has discretized the system in a way that preserves the gauge symmetries, and there is a good control over the gauge invariant observables (i.e.~Wilson loops). Real space renormalization is being carried out in order to f\/ind the UV completion of the theory, i.e.~incorporate all degrees of freedom by performing the continuum limit. The method of perfect actions is also applied here, but is usually performed to minimize lattice artifacts, and in order to restore the space-time symmetries which have been broken by the lattice, e.g.~rotations. Also, they are a tool in order to improve the convergence to the continuum limit when taking the lattice spacing $a\to 0$ \cite{Hasenfratz:1997ft}.

In the case of GR however, the gauge symmetry and the dynamics are intimately interwined. Being on the gauge-invariant sector of the theory implies being on the solution space of GR as well. Hence a discrete approximation will only have exact gauge symmetry if it contains all the information of the continuum theory, and therefore sits at the RG f\/ixed point.

This adds another point of view to the issue of renormalization in lattice theories to the case of discrete gravity. On the one hand, the RG f\/low can just be seen as a method of investigating the behavior of a theory described by a (fundamental) lattice model, on dif\/ferent scales. In particular, one can ask under which circumstances the UV limit of the theory describes GR, i.e.~in which sense dif\/feomorphism symmetry emerges. In this case, one does not necessarily have to demand dif\/feomorphism invariance of the discrete theory. As a result, the discrete theory will contain information about more degrees of freedom than just pure GR. These pseudo-degrees of freedom arise in the same way in which e.g.~in electromagnetism breaking of the gauge symmetries would result in the emergence of longitudinal photons. In this framework, the main contribution of the path integral would come from triangulations with a large number of vertices, which would render the expansion into vertex numbers problematic \cite{Henderson:2010qd}.

From this point of view, the investigation of Spin Foam renormalization deals with the question of how the pseudo degrees of freedom decouple in the thermodynamic limit\footnote{Note that this is also connected to the question of whether in this limit the path integral is dominated by states which resemble smooth manifolds. This is a long-standing question also in other approaches to discrete gravity such as causal dynamical triangulations~\cite{Ambjorn:2011ph}, and is also under intensive investigation in the GFT approach to spin foams~\cite{Bonzom:2011zz}.}.

On the other hand, if one were to strictly demand the correct implementation of dif\/feo\-mor\-phism symmetry at all scales, then one would have to demand the gauge symmetry to be present on the discrete level (i.e.~on a triangulation). One would then have to construct the path integral measures with the correct symmetries, and the analysis of the last sections suggest that renormalization techniques are a helpful tool to achieve this. Only in this case would the corresponding canonical theory support the constraints satisfying the Dirac algebra, exactly also for coarse triangulations. In particular, the theory would most likely be independent under Pachner moves, as has been argued in \cite{Barrett:1995mg, Pfeiffer:2003tx}.

Both points of view are physically quite dif\/ferent, and the resulting theories for either case would only agree for large triangulations. The investigation of both, however, require the use of renormalization group techniques, albeit for dif\/ferent purposes.

To investigate these issues, a collection of toy models has been introduced in \cite{Bahr:2011yc}, where the gauge groups within spin foam models are replaced by f\/inite groups, making them more accessible to numerical studies and avoiding many issues with divergencies. In \cite{Dittrich:2011zh} the coarse graining methods described above have been applied to a class of so-called cut-of\/f models based on the Abelian group $\mathbb{Z}_p$. In order to deal with the non-local terms which are prevalent in real space renormalization approaches, a Migdal--Kadanof\/f-type approximation scheme has been employed \cite{Migdal:1975zg}. Already in $3D$ one can see that it is a highly nontrivial issue whether the discrete model f\/lows to a f\/ixed point at which dif\/feomorphism symmetry is realized. The behavior of RG trajectories  depends highly on the starting point, leading to a structured phase diagram with several dif\/ferent f\/ixed points.

\section{Summary}

In this article, we have reviewed various angles of the problems that arise in discretizing theories with dif\/feomorphism symmetry, with (quantum) general relativity in mind.

On the one side, we have discussed in Section~\ref{Ch:Consistent} how the constraint algebra can be deformed, and how this can lead to all sorts of undesirable problems, such as Lagrange multipliers becoming complex, and one essentially losing control over the quality of the approximation that the discrete theory provides for the continuum theory. To remedy this problem, the uniform discretization approach was presented, which can be seen as a concrete implementation of the master constraint programme. Although the symmetries are not restored on the discrete level, one has good control over the value of the constraints, which can be made arbitrarily small, and constant over time. We have treated two examples, one with a good continuum limit, and another one with a~fundamental level of discreteness, in which the continuum theory is only approximated well in some sense. It is widely believed that this should be a characteristic of a quantum theory of gravity.

On the other side, in Section~\ref{Sec:ImprovedAndPerfectActions} we have demonstrated in which sense the dif\/feomorphism symmetry is broken in the covariant setting, in particular in Regge Calculus. We have described how one might hope to restore dif\/feo symmetry on the discrete level, by replacing the Regge action with a so-called perfect action, which can be def\/ined by an iterative coarse-graining process. On the quantum side, this process resembles a Wilsonian renormalization group f\/low, and it has been shown how, with this method, one can construct both the classical discrete action, as well was the quantum mechanical propagator, with the correct implementation of dif\/feomorphism symmetry, in certain mechanical toy models, as well as for $3d$ GR. Whether a similar construction works in $4d$ general relativity is still an open question, being related to issues of locality and renormalizability, which are still largely open in this context.

\pdfbookmark[1]{References}{ref}
\LastPageEnding

\end{document}